\documentclass[aps,prl,showpacs,preprintnumbers,twocolumn,superscriptaddress]{revtex4-2}
\usepackage{amsmath,amssymb}
\usepackage{bm}
\usepackage{tipa}
\usepackage{upgreek}
\usepackage{comment}
\usepackage{mathrsfs}
\usepackage{graphicx}
\usepackage{braket}
\usepackage{enumitem}
\usepackage{mathbbol}
\usepackage{booktabs}
\usepackage{gensymb}
\usepackage[normalem]{ulem}
\usepackage{color}
\usepackage[colorlinks,bookmarks=true,citecolor=blue,linkcolor=red,urlcolor=blue]{hyperref}
\usepackage{hyperref}
\renewcommand{\vec}[1]{\mathbf{#1}}
\usepackage{pifont}

\begin{document}

	\title{Interaction effects in a 1D flat band at a topological crystalline step edge} 
	
	\author{Glenn Wagner}
	\altaffiliation{These authors contributed equally to this work.}
	\affiliation{Department of Physics, University of Zurich, Winterthurerstrasse 190, 8057 Zurich, Switzerland}
	\author{Souvik Das}
	\altaffiliation{These authors contributed equally to this work.}
	\affiliation{Max Planck Institute of Microstructure Physics, Halle 06120, Germany}
	\author{Johannes Jung}
	\affiliation{Physikalisches Institut, Experimentelle Physik II, Universität Würzburg, Am Hubland, 97074 Würzburg, Germany}
	\author{Artem Odobesko}
	\altaffiliation{Corresponding author}
	\affiliation{Physikalisches Institut, Experimentelle Physik II, Universität Würzburg, Am Hubland, 97074 Würzburg, Germany}
	\author{Felix Küster}
	\affiliation{Max Planck Institute of Microstructure Physics, Halle 06120, Germany}
	\author{Florian Keller}
	\affiliation{Physikalisches Institut, Experimentelle Physik II, Universität Würzburg, Am Hubland, 97074 Würzburg, Germany}
	\author{Jedrzej Korczak}
	\affiliation{Institute of Physics, Polish Academy of Sciences, Aleja Lotnik\'ow 32/46, 02-668 Warsaw, Poland}
	\affiliation{International Research Centre MagTop, Institute of Physics, Polish Academy of Sciences, Aleja Lotnik\'ow 32/46, 02-668 Warsaw, Poland}
	\author{Andrzej Szczerbakow}
	\affiliation{Institute of Physics, Polish Academy of Sciences, Aleja Lotnik\'ow 32/46, 02-668 Warsaw, Poland}
	\author{Tomasz Story}
	\affiliation{Institute of Physics, Polish Academy of Sciences, Aleja Lotnik\'ow 32/46, 02-668 Warsaw, Poland}
	\affiliation{International Research Centre MagTop, Institute of Physics, Polish Academy of Sciences, Aleja Lotnik\'ow 32/46, 02-668 Warsaw, Poland}
	\author{Stuart Parkin}
	\affiliation{Max Planck Institute of Microstructure Physics, Halle 06120, Germany}
	\author{Ronny Thomale}
	\affiliation{Institut für Theoretische Physik und Astrophysik Universität Würzburg, 97074 Würzburg, Germany}
	\author{Titus Neupert}
	\affiliation{Department of Physics, University of Zurich, Winterthurerstrasse 190, 8057 Zurich, Switzerland}
	\author{Matthias Bode}
	\affiliation{Physikalisches Institut, Experimentelle Physik II, Universität Würzburg, Am Hubland, 97074 Würzburg, Germany}
	\author{Paolo Sessi}
	\affiliation{Max Planck Institute of Microstructure Physics, Halle 06120, Germany}
	
	\begin{abstract}
    Step edges of topological crystalline insulators can be viewed as predecessors of higher-order topology, as they embody one-dimensional edge channels embedded in an effective three-dimensional electronic vacuum emanating from the topological crystalline insulator. Using scanning tunneling microscopy and spectroscopy we investigate the behaviour of such edge channels in Pb$_{1-x}$Sn$_{x}$Se under doping.  Once the energy position of the step edge is brought close to the Fermi level, we observe the opening of a correlation gap. The experimental results are rationalized in terms of interaction effects which are enhanced since the electronic density is collapsed to a one-dimensional channel. This constitutes a unique system to study how topology and many-body electronic effects intertwine, which we model theoretically through a Hartree-Fock analysis.
	\end{abstract}
	
 	\maketitle

\textbf{Introduction ---} The hallmark feature of three-dimensional topological insulators (TIs)~\cite{TIs,Qi} are their protected gapless surface states with the dispersion of an \emph{odd} number of massless Dirac fermion. These surface states have a property called \emph{chirality}, which makes them anomalous: It is not possible to obtain these two-dimensional surface states without incorporating the three-dimensional bulk. Mathematically, this is encoded in the fermion doubling theorem~\cite{NN1,NN2,NN3} which says that it is not possible to obtain fermions of a single chirality in a purely two-dimensional system with time-reversal. 	

\begin{figure*}
    \centering
    \includegraphics[width=1\textwidth,page=1]{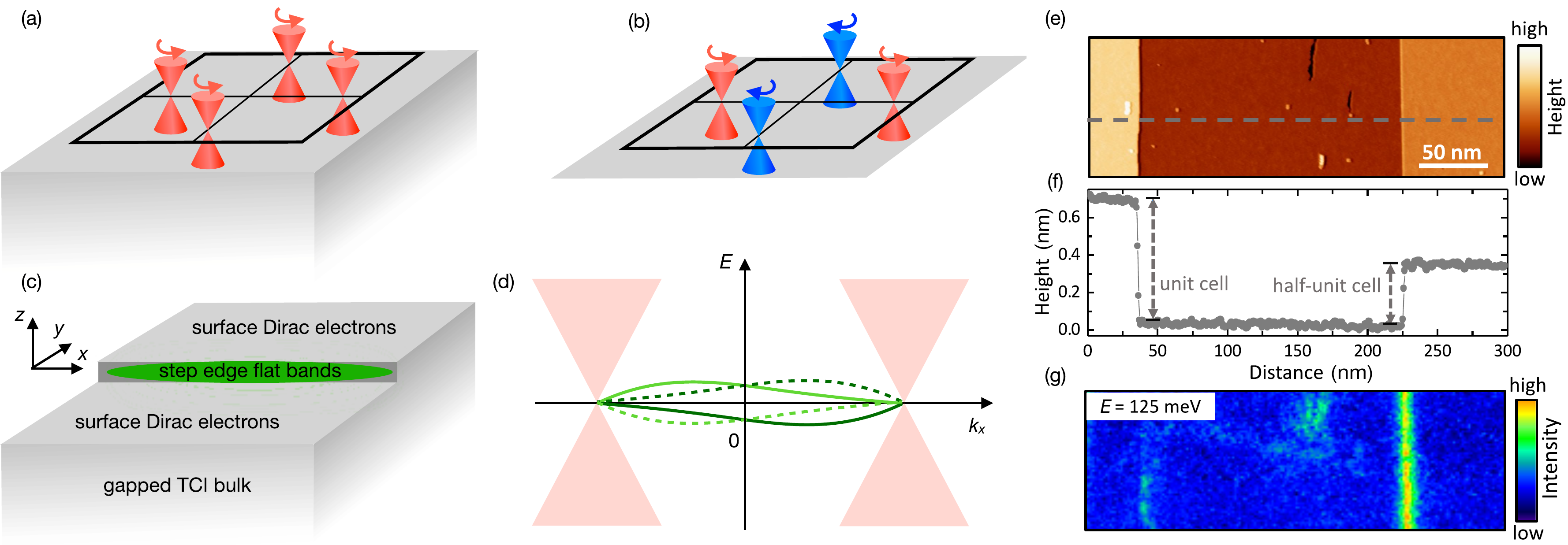}
    \caption{\textbf{Emergence of 1D flat bands in TCI.} (a) The TCI Pb$_{1-x}$Sn$_{x}$Se has four Dirac cones of the same chirality in the BZ. (b) For a purely 2D system there would be an equal number of Dirac cones with positive and negative chirality. (c) 1D flat bands emerge at a step edge on the TCI surface. (d) Band structure Eq.~\eqref{eq:kinE} of the four edge mode states along with the surface Dirac cones. (e) STM topographic image acquired at the (001) surface of pristine  Pb$_{0.7}$Sn$_{0.3}$Se. The dashed  grey line corresponds to the line profile reported in (f). Two different steps are visible, corresponding to unit and half-unit cell heights. (g) $dI/dU$ map acquired at the Dirac point ($E_{\textrm{D}} =$ + 125 meV). The signal, proportional to the sample local density of states, shows a strong enhancement localized around the half unit cell step. Scanning parameters: $V = 125$ mV, $I = 250$ pA, $V_{\rm rms} = 10$ meV. }
    \label{fig:ExpOne}
\end{figure*}

Topological crystalline insulators (TCI) are TIs that are protected by crystalline symmetries \cite{TCI,Ando}. In contrast to TIs, the surface of this TCI can host \emph{multiple} Dirac cones, which are all of the \emph{same} chirality (see Fig.~\ref{fig:ExpOne}a), and exhibit the rotation anomaly: A purely two-dimensional model would have an equal number of Dirac cones with positive and negative chirality (see Fig.~\ref{fig:ExpOne}b) \cite{RotAnomaly}. In this work we investigate the one-dimensional edge states arising at odd-atomic step edges on the surface of the TCI Pb$_{1-x}$Sn$_x$Se (Fig.~\ref{fig:ExpOne}c).  The detection of these spin-polarized midgap states at step edges on the surface of Pb$_{1-x}$Sn$_x$Se was described in previous work including some of the present authors~\cite{Sessi}, which was confirmed in Ref.~\onlinecite{Madhavan} and further theoretically detailed in Ref.~\onlinecite{Buczko}. In this contribution, we report scanning tunneling microscopy (STM) and spectroscopy (STS) measurements of these edge states using surface doping to controllably tune their energy position with respect to the Fermi level.  This experimental approach is used to systematically scrutinize the emergence of correlation effects under the effect of distinct dopants.

In typical 3D TIs the Coulomb interaction is not strong enough to lead to spontaneous symmetry breaking in the two-dimensional surface states \cite{Stern}. For Pb$_{1-x}$Sn$_x$Se with its large dielectric constant which effectively screens electron-electron interactions, correlation effects are generally disregarded \cite{Nimtz}. However, the 1D flat bands, which reside at step edges, are characterized by an enhanced density of states which can lead to correlated states (Fig.~\ref{fig:ExpOne}d). For example, in an attempt to provide a possible explanation for the zero-bias
conductance peak observed in the point contact spectroscopy experiments \cite{PhysRevB.100.041408}, it has been suggested that 1D flat bands might be susceptible to correlation-driven instabilities resulting in the formation of magnetic domains \cite{PhysRevB.100.121107}. Similar flat boundary states are known to arise in a variety of systems, such as graphene \cite{Ryu}, topological semimetals \cite{Chan} and $d$-wave superconductors \cite{Wang_FeSC}, which in some cases exhibit spontaneous symmetry breaking. In the present case, the edge modes have a flat dispersion and are therefore susceptible to flat-band Stoner ferromagnetism---a one-dimensional analogue of quantum Hall ferromagnetism in the zeroth Landau level (LL) of graphene~\cite{QHFM_graphene,Young2012} or in twisted bilayer graphene~\cite{GroundStateHiddenSymmetry,TBG_FM}. 

Spontaneous symmetry breaking is associated with the opening of correlation gaps. Our spectroscopic measurements reveal two different behaviours depending on the position along the step edge where the measurement is taken. When the energy of the 1D flat band  is tuned to the Fermi level, the single peak in the density of states (DOS) from the edge mode either splits in two or four peaks. We explain this behaviour theoretically in terms of different states that spontaneously break time-reversal symmetry.

\textbf{Experiments ---} Pb$_{1-x}$Sn$_{x}$Se crystallizes in rock salt structure for $x  \leq 0.4$. Previous studies showed how this compound can host two topological distinct phases \cite{Hsieh}. Starting from PbSe, a trivial narrow band gap semiconductor, the system undergoes a topological phase transition by progressively increasing the Sn concentration. At low temperature, the topological crystalline phase is observed for $x  \geq 0.2$ \cite{Dziawa}. In the present study, we focus on Pb$_{0.7}$Sn$_{0.3}$Se single crystals grown by the self-selecting vapor method \cite{Dziawa,Sessi}.  Our crystals are thus safely inside the topological crystalline regime of the Pb$_{1-x}$Sn$_{x}$Se phase diagram. Single crystals have been cleaved at room temperature in ultra-high vacuum conditions ($p < 5\cdot10^{-10}$ mbar). Experiments have been performed in two distinct STM set-ups, operated at $T$ = 2 K and $T$ = 4.5 K. All measurements have been acquired using electro-chemically etched tungsten tips. Differential conductance $dI/dU$ data have been measured by lock-in technique by applying a bias voltage modulation $V_{\rm rms}$ to the tip.

\begin{figure*}
    \centering
    \includegraphics[width=\textwidth,page=1]{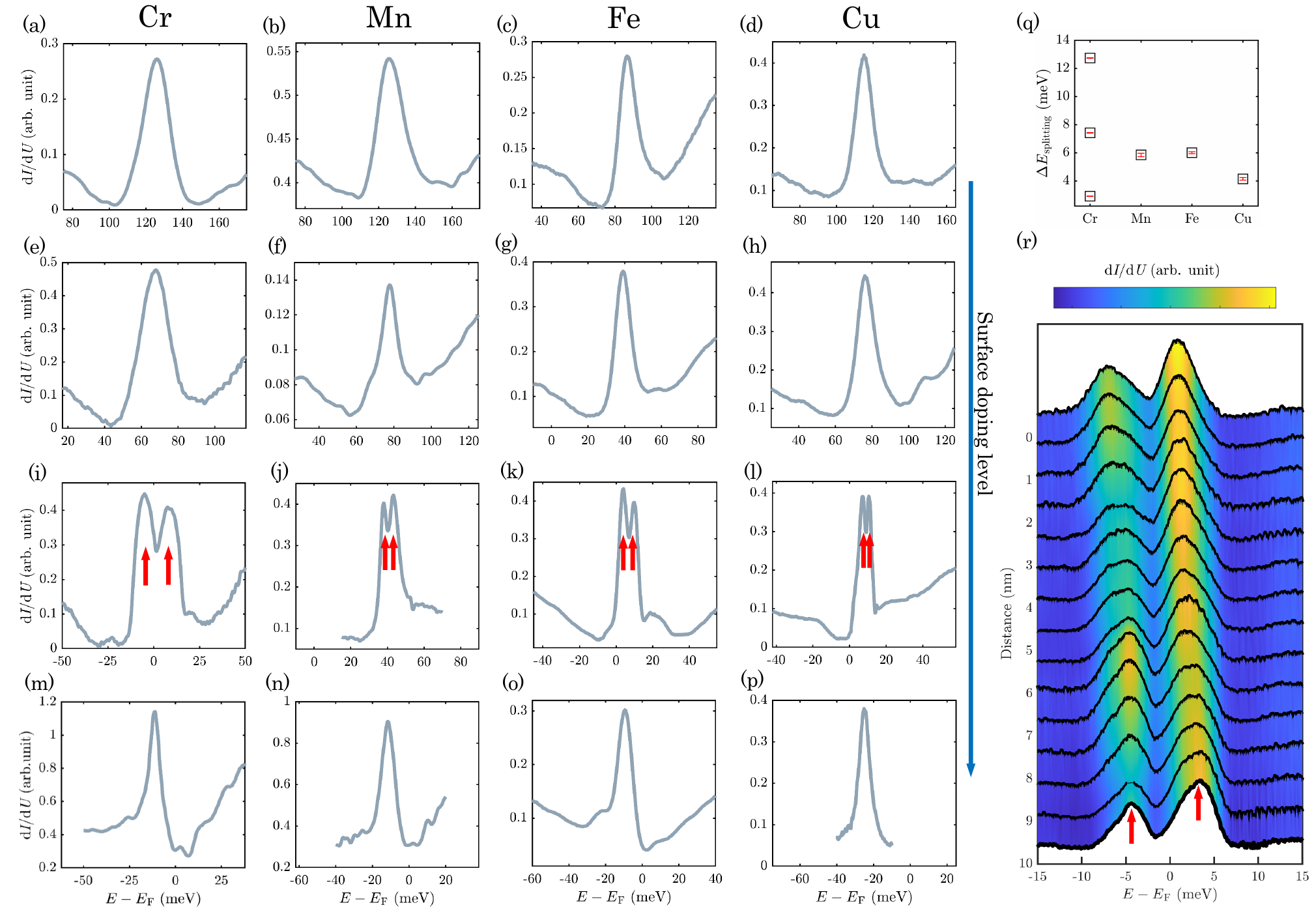}
    \caption{\textbf{Doping dependence of the 1D flat band} (a-p) Scanning tunnelling spectroscopy of the 1D flat band emerging at half unit cell steps as a function of the doping level. Each column reports the energy evolution of the 1D flat band at different doping concentrations for distinct dopants (Cr, Mn, Fe, and Cu). Measurements on Cr- and Mn-doped samples have been performed at $T$ = 2 K. Measurements on Fe- and Cu-doped samples have been performed in a different set-up operated at $T$ = 4.5 K. For all elements, the deposition onto the Pb$_{0.7}$Sn$_{0.3}$Se surface provides a $n$-doping effect. Starting from $p$-doped crystals, this procedure allows to progressively shift the energy of the 1D flat band towards the Fermi level. A splitting of the local density of states into a double peak structure is visible in (i-l), signaling the emergence of interaction effects. By continuosly doping the surface, the splitting disappears once the 1D flat band is shifted below the Fermi level. (q) Magnitude of the splitting observed in panels (i-j). (r) Line spectroscopy acquired along the step edge once the energy of the 1D flat band is brought close to the Fermi level (Cr adatoms as dopants). The spectra evidence the existence of spatial fluctuations of the double peak structure, an effect attributed to the disorder created by the random distribution of dopants.}
    \label{fig:EXP_combined}
\end{figure*}

Figure~\ref{fig:ExpOne}e shows a STM topographic image acquired in constant-current mode on a freshly cleaved Pb$_{0.7}$Sn$_{0.3}$Se crystal. The exposed surface corresponds to the (001) orientation which is commonly obtained when cleaving a bulk crystal \cite{Dziawa,Tanaka,Xu,Okada,PhysRevLett.126.236402}. At this surface, angle-resolved photoemission studies revealed the presence of four Dirac cones protected by mirror symmetry located close to the $\overline{X}$ and $\overline{Y}$ points of the Brillouin zone \cite{Dziawa,Tanaka,Xu,Polley2018}. The topographic image shows large terraces separated by step edges which, as highlighted by the line profile reported in Fig.~\ref{fig:ExpOne}f, are characterized by different heights. These two steps are representative of two distinct classes, namely, (i) steps whose height is equal to an integer multiple of the lattice constant  $n \cdot a$, and (ii) steps whose height is a half-integer multiple of the lattice constant $(1/2 + n) a$ with  $n$ being the integer and $a$ the lattice constant ($a~\approx$~6\,\AA). As described in Ref.~\onlinecite{Sessi}, while the translation symmetry of the surface lattice is preserved for integer multiple steps, half-integer multiple steps introduce a 1D structural $\pi$-shift which dramatically influences the surface electronic properties. This is illustrated in Fig.~\ref{fig:ExpOne}g, which reports a $dI/dU$ map acquired at the Dirac point located at $E_{\textrm{D}} \approx$ + 125 meV (see Supplementary Figure~\ref{fig:SuppSTS} for a description of the energy level alignment). The $dI/dU$ signal, which is proportional to the sample local density of states, shows a strong enhancement  at the half-integer step. As discussed in  Ref.~\onlinecite{Sessi} and Figure~\ref{fig:SuppSTS}, this corresponds to the spectroscopic signature of a 1D flat band localized around the 1D structural $\pi$-shift.

The present system thus represents an ideal platform to scrutinize the emergence of interaction effects in 1D flat bands, which are expected to manifest once the flat bands are energetically localized close to the Fermi level. The key idea is that, as the kinetic energy is quenched, electron correlations can become the dominant energy scale. 
To experimentally realize such a scenario, the 1D flat band has to be tuned to the Fermi level. To achieve this goal, we used a surface doping approach. Starting from pristine $p$-doped crystals ($E_{\textrm{D}}$ in the range +90--125 meV, see Fig.~\ref{fig:EXP_combined}a--d), we progressively dose higher amounts of distinct 3$d$ adatoms onto the crystal surface held at cryogenic temperature, a procedure known to create a downwards band bending, i.e.\ a rigid shift towards negative energies \cite{SRB2014}.

Figure~\ref{fig:EXP_combined}a--p summarizes the spectroscopic results as a function of the doping level, with each column corresponding to a distinct dopant, namely Cr, Mn, Fe, and Cu. Starting from pristine samples (see Fig.~\ref{fig:EXP_combined}a--d), a rigid shift towards negative energy is observed upon deposition onto the surface, irrespective of which specific 3d element is used, as illustrated in Fig.~\ref{fig:EXP_combined}e--h. The shift successively increases with each deposition step. Once the 1D flat band hits the Fermi level, the single peak corresponding to its spectroscopic signature is found to split into a double peak structure, as highlighted by the red arrows in Fig.~\ref{fig:EXP_combined}i--l. Occasionally, a splitting of the 1D flat band is not only observed if $E_{\rm D}$ reaches the Fermi level but already in its vicinity, as evidenced in Fig.~\ref{fig:EXP_combined}j. A similar splitting has been previously reported in the 2D case for highly degenerate Landau levels in graphene and attributed to broken valley degeneracy \cite{Song}. By further increasing the concentration of surface dopants, the Dirac point is shifted below the Fermi level which results in the recovery of a single peak structure characteristic of the 1D flat band, as illustrated in Fig.~\ref{fig:EXP_combined}m--p.

In all cases, the size of the splitting amounts to a few meVs, as summarized in Fig.~\ref{fig:EXP_combined}q. The different data points reported for Cr correspond to distinct experimental runs. They reveal a distribution in the size of the splitting which is not linked to the specific element, but which is attributed to the combined effect of (i) intrinsic samples inhomogeneities and (ii) the disorder induced by the random distribution of the dopants. This is demonstrated by the additional spectroscopic data reported in Fig.~\ref{fig:SuppSplit}. Although starting from a nominally equivalent sample, i.e. Pb$_{0.7}$Sn$_{0.3}$Se with Dirac point located at ~125 meV, the splitting detected once the 1D flat bands hits the Fermi is lower (3 meV) with respect to the one reported in Fig.~\ref{fig:EXP_combined}i (12.5 meV). A direct comparison of spectroscopic and topographic images suggests the size of the splitting to be inversely proportional to the amount of disorder, as illustrated in Fig.~\ref{fig:Disorder}.

To test the robustness of this observation against potential artifacts, we performed numerous control experiments.  For example, in order to exclude an uncontrolled influence of a spatial inhomogeneity of the TCI surface, the very same sample region was mapped before and after deposition, as illustrated in Supplementary Figure \ref{fig:samelocation}.  Moreover, we verified that integer step edges under the same doping conditions, i.e.\ once the Dirac point is tuned to the Fermi, do not show any significant change with respect to the spectral shape observed in the pristine case, see Supplementary Figure \ref{fig:integerstep}.  This ensures that the observed behaviour is indeed linked to the evolution of the electronic properties of the 1D flat band hosted at half-integer steps as a function of doping level. 

Note that, although this surface doping approach allows to controllably shift the Dirac point towards the Fermi level, the random distribution of dopants inevitably increases the surface disorder after each deposition step. This results in spatial fluctuations of the Dirac point illustrated in Fig.~\ref{fig:EXP_combined}q, which reports spatially resolved scanning tunneling spectroscopy acquired at distinct positions along a structural $\pi$ shift. Although these  data evidence the existence of different broadening as well as fluctuations in the peak intensity, the peak splitting remains clearly visible along the entire profile.

\begin{figure}
    \centering
    \includegraphics[width=0.3\textwidth,page=1]{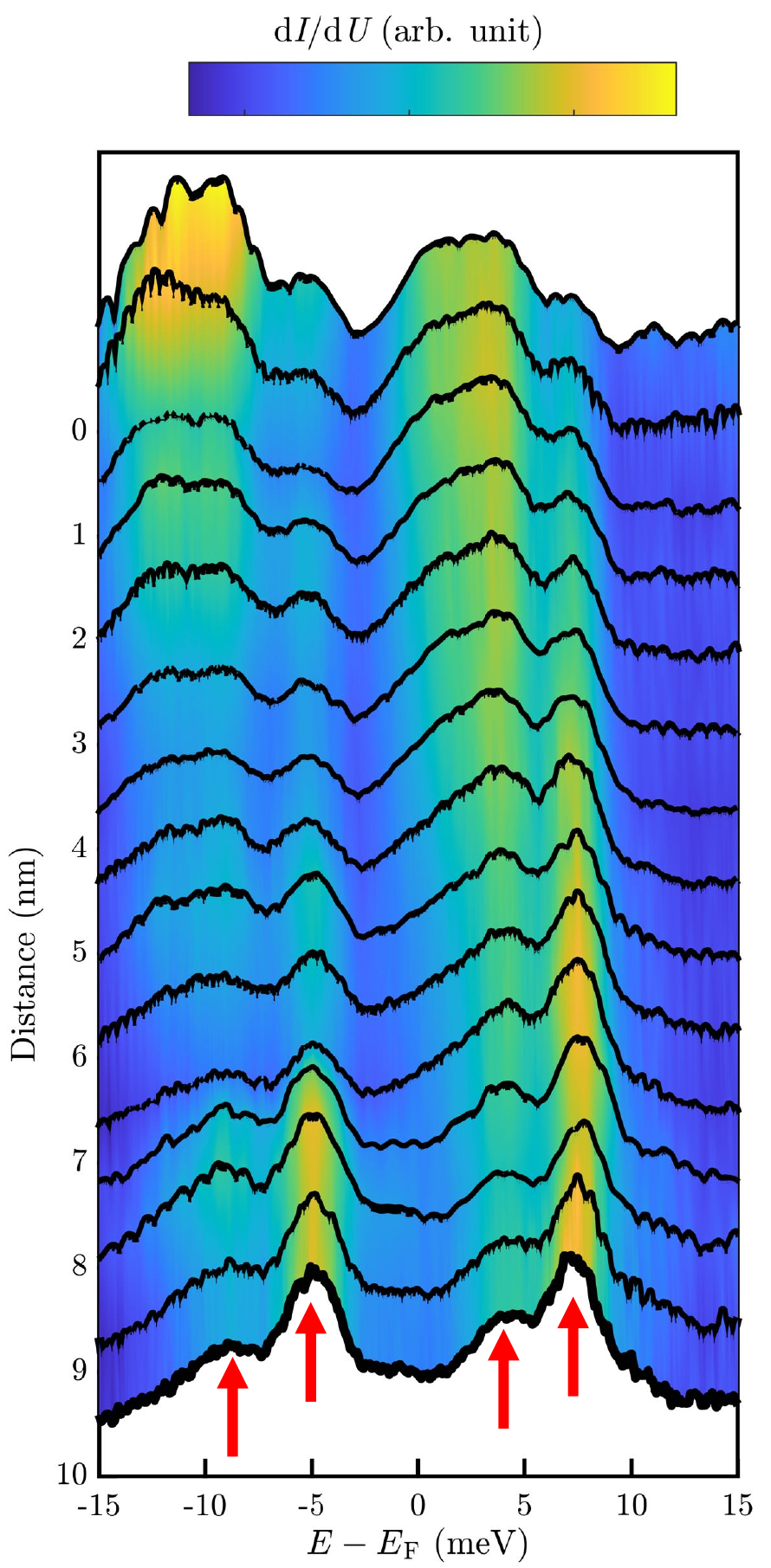}
    \caption{\textbf{Spectroscopic signatures of interaction effects}. Scanning tunneling spectroscopy data acquired along a step corresponding to a structural $\pi$ shift once the Dirac point is located at the Fermi level. Cr adatoms have been used as dopants. A multi-peaks splitting of the local density of states associated to the flat band is visible. This corresponds to a four-peaks structure (see red arrows) which, as in the double peak case, is characterized by spatial fluctuations induced by the disorder.}
    \label{fig:EXP_four_peaks}
\end{figure}

A splitting of the 1D flat band into a double-peak structure is predominantly found in our samples once the Dirac point hits the Fermi level. However, our measurements occasionally reveal the existence of a spectroscopically more rich scenario where the local density of states associated to the 1D flat band splits into a multi-peak structure. This effect is illustrated in the spatially resolved spectroscopic data reported in Fig.~\ref{fig:EXP_four_peaks} for Cr adatoms, i.e. the dopant most systematically used in our study. They show a four-peak structure (see red arrows) which, as for the double-peak case discussed in Fig.~\ref{fig:EXP_combined}q, is characterized by spatial fluctuations induced by the disorder. As discuss in the theory section, this effect is in agreement with our theoretical analysis, being a direct fingerprint of two distinct energy scales.

\begin{figure}
    \centering
    \includegraphics[width=0.4\textwidth]{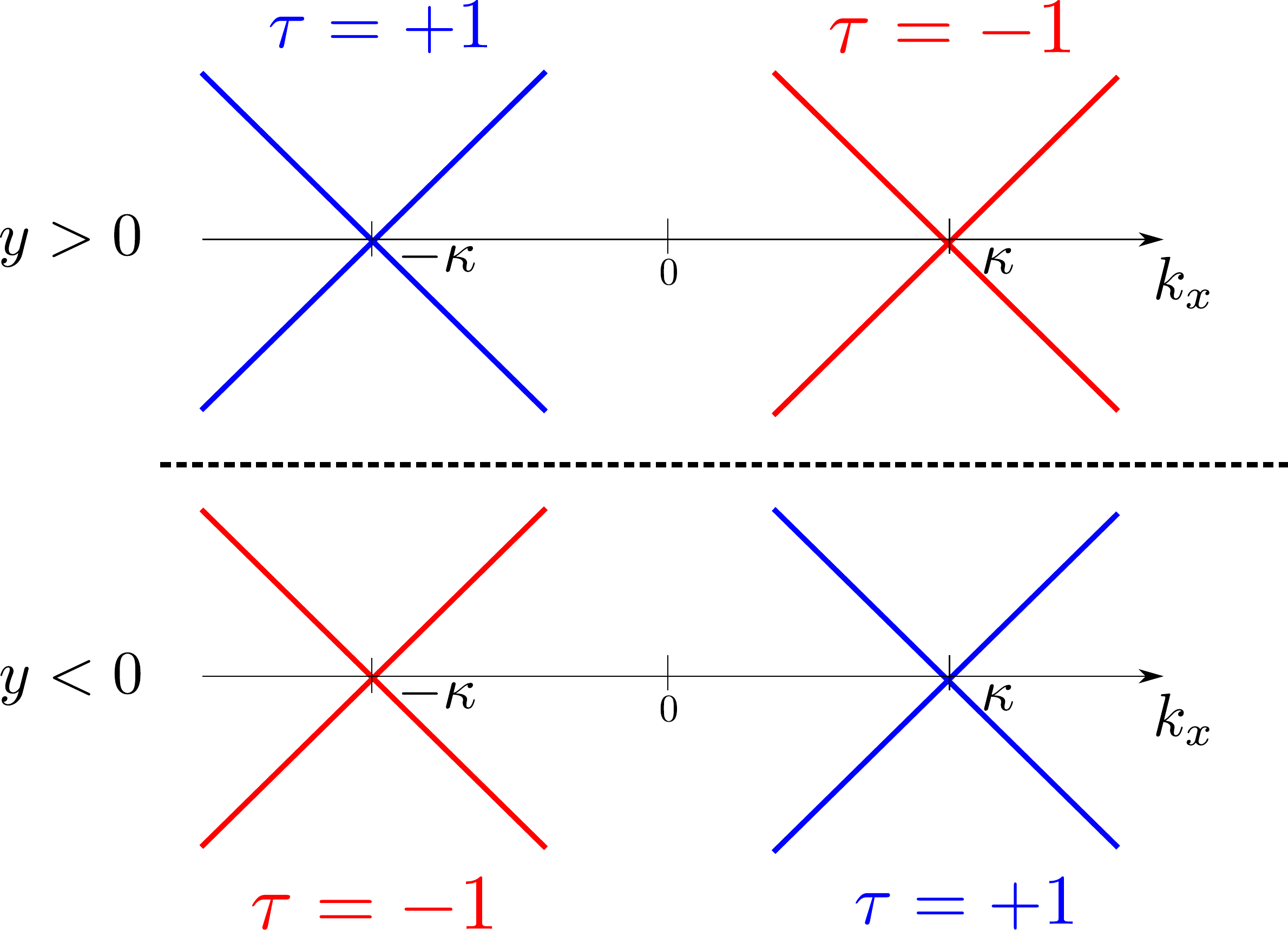}
    \caption{Schematic of the toy model Eq.~\eqref{eq:toy_model} we consider for the step edge. We shift the two Dirac cones of the two valleys by an amount $\kappa$ along the $x$-axis. We take this shift to be opposite for $y>0$ and $y<0$, this way we obtain an edge mode at $y=0$.}
    \label{fig:setup}
\end{figure}

\textbf{Theory ---} The $k\cdot p$ theory for Pb$_{1-x}$Sn$_{x}$Se has been worked out in Refs.~\onlinecite{k_dot_p,Wang_SnTe} and the corresponding Landau level spectrum was discussed in Ref.~\onlinecite{LLspectrum}. Here, as a model we propose a more simple Hamiltonian consisting of four Dirac points in the BZ at $(\pm\kappa,\pm\kappa)$:
\begin{equation}
H=v_{\rm F}\left[(p_{y}-\kappa_y) \sigma_{x}-(p_{x}-\kappa_x) \sigma_{y}\right] ,
\label{eq:toy_model}
\end{equation}
where $p_i=-i\partial_i$, and $\sigma_i$ are the Pauli matrices associated with spin. We label the valleys by two pseudospin degrees of freedom $\tau_i,\eta_i$. The step edge is manifest as an exchange of the valleys between $y>0$ and $y<0$, such that $y=0$ is the location of the step edge (see Fig.~\ref{fig:setup}): $(\kappa_x,\kappa_y)=\kappa(\tau_z\textrm{sign}(y),\eta_z)$. Estimates for the Fermi velocity $v_{\rm F}$ can be found in Refs.~\onlinecite{Tikui,Liang2013}.

We label the eigenstates by their $\sigma_z,\tau_z,\eta_z$ eigenvalues $\sigma,\tau,\eta$. There are four zero modes in the range $-\kappa<k_x<\kappa$ which are localized around $y=0$ with opposite spins in the two valleys (i.e.~the eigenvalue of $\sigma_z$ is the same as the eigenvalue of $\tau_z$ \footnote{This is analogous to the lowest Landau level of a chiral Dirac fermion being spin-polarized.}):
\begin{equation}
\psi_{\tau\eta k_x}(x,y)=e^{ik_xx+i\kappa\eta y}\begin{cases}
e^{-|k_x-\tau\kappa|y}, & y>0\\
e^{|k_x+\tau\kappa|y}, & y<0
\end{cases}.
\label{eq:psi_zero_B}
\end{equation}

We study the symmetry breaking patterns in the edge states due to electron-electron interactions. This problem is reminiscent of the long-standing problem of magnetism in graphene edges. The zig-zag edge of graphene hosts an exact zero energy mode \cite{Fujita,Nakada} (a finite dispersion for the edge modes can be generated by next-nearest neighbour hopping) and interactions lead to a ferromagnetic state, as shown in Hartree-Fock \cite{Fujita}, exact diagonalization \cite{Guinea}, perturbative approaches \cite{Affleck}, and bosonization \cite{Schmidt}. Furthermore, in graphene nanoribbons, the two edges can be coupled by interactions, leading to antiferromagnetic inter-edge coupling \cite{Lieb,Fujita,castro2008magnetic}. By a similar mechanism, the Majorana flat bands in $d$-wave superconductors order magnetically \cite{Potter}. We choose to study the step-edge problem in a similar vein and rely on the Hartree-Fock approximation, since for zig-zag edges more sophisticated techniques yield similar results. There are two important differences between the zig-zag edges of graphene and the step-edge modes studied here. Firstly, we have twice the number of flat bands, namely four instead of two. Secondly, unlike in graphene the edge modes in the TCI are not spin-degenerate since Pb$_{1-x}$Sn$_{x}$Se exhibits a significant spin-orbit coupling. 

In the phenomenological model introduced above, we obtain fully flat bands for the edge states. However, a microscopic three-dimensional model finds edge states with a finite dispersion~\cite{Sessi}, hence we add this dispersion by hand. The bands calculated in Ref.~\onlinecite{Sessi} have two van Hove singularities (VHS). One of the VHS arises where the flat band merges with the Dirac cone, at which point the states also get more extended perpendicular to the edge. Therefore we expect only the other VHS to show up as a peak in the edge density of states (DOS) measured by the STS. This motivates the following model for the dispersion (Fig.~\ref{fig:ExpOne}d):
\begin{equation}
    \epsilon_{k_x\tau\eta}=W\eta\bigg(\tau\cos\frac{\pi k_x}{2\kappa}+\frac{1}{5}\sin\frac{\pi k_x}{\kappa}\bigg),
    \label{eq:kinE}
\end{equation}
with $W$ the bandwidth. The full second-quantized Hamiltonian is of the form $H = H_\textrm{kin} + H_\textrm{int}$ where $H_\textrm{kin}=\sum_{\alpha}\epsilon_\alpha c_{\alpha}^\dagger c_{\alpha}$ and the interaction term will be
\begin{equation}
   H_\textrm{int}= \frac{1}{2}V_{\alpha\beta\gamma\delta}c^\dagger_\alpha c^\dagger_\beta c_\delta c_\gamma,
\end{equation}
where the use the short-hand label $\alpha=(k_x,\tau,\eta)$. The matrix elements $V_{\alpha\beta\gamma\delta}$ are obtained by projecting the Coulomb interaction onto the flat bands. Since our model is purely a two-dimensional model of the surface, we use the two-dimensional Coulomb interaction $V_q=\frac{e^2}{2\epsilon_0q}$. Screening from the three-dimensional bulk may result in a renormalized dielectric constant. We perform a mean-field decoupling of the Hamiltonian and solve the Hartree-Fock equations self consistently (see Supplement \cite{Supplement} for details).

\begin{figure}
    \centering
    \includegraphics[width=\columnwidth]{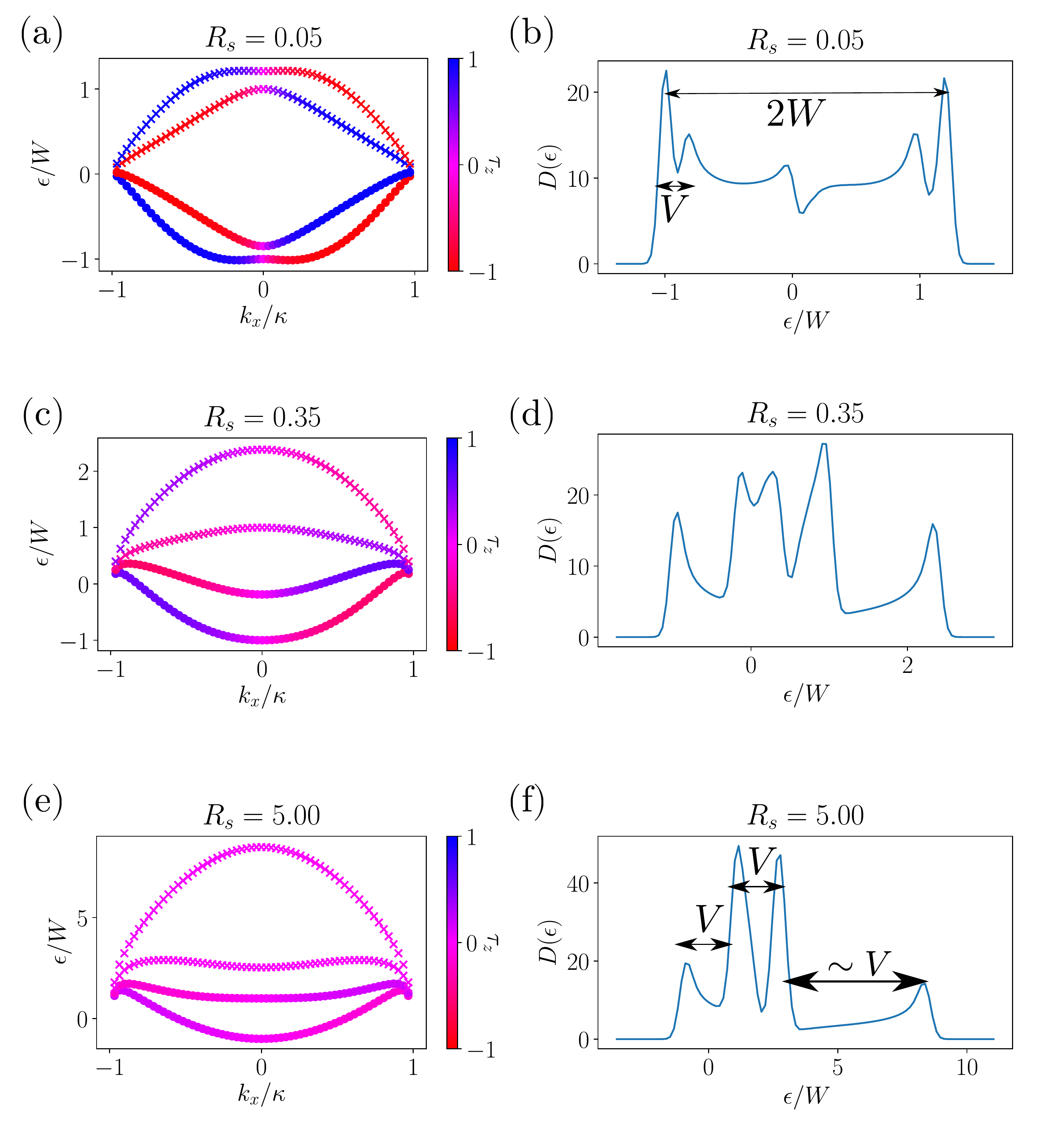}
    \caption{\textbf{Hartree-Fock results} The HF bandstructures (left) along with the corresponding DOS (right) are shown for three values of $R_s$ and $\bar\kappa=1$. The colorbar in the bandstructure shows $\langle\tau_z\rangle$.  As $R_s$ increases, we obtain a two and four peak structure. The conduction and valence band are split by the energy scale $W$. Hybridization between orbitals leads to interaction-induced gaps of order $V$ opening up. The circles indicate filled states while the crosses indicate empty states. We pick a resolution of $N_y=41$ for $k_y$.}
    \label{fig:HF}
\end{figure}

There are two energy scales in the problem. The kinetic energy scale is the bandwidth $W$, while the interaction energy scale is $V= \frac{e^2\kappa}{2\epsilon_0}$. The model thus has two dimensionless parameters, $\bar\kappa=\kappa a$ ($a$ is the lattice spacing in the $y$-direction) and $R_s=\frac{V}{W}$. The qualitative results are largely independent of $\bar\kappa$; for the bandstructure of the TCI in question in this work, we have $\bar\kappa=0.5$~\cite{Sessi}. Rather, we focus on the dependence on $R_s$. Let us consider this model at half filling. The HF results are shown in Fig.~\ref{fig:HF}. In the limit $R_s\ll 1$ we completely fill the valence band subspace ($\eta=-1$) and the interaction leads to a slight hybridization between the opposite spin bands at the band crossing (Fig.~\ref{fig:HF}a). This state spontaneously breaks time-reversal symmetry and leads to two peaks in the DOS (Fig.~\ref{fig:HF}b), with a splitting given by $W$. In the limit $R_s\sim 1$ the splitting between the conduction and valence band ($\sim W)$ remains, and the valence band subspace is completely filled. Due to the interaction, however, the opposite spin bands in the valence and conduction band subspaces are fully hybridized forming bonding and anti-bonding orbitals, which are split by an amount $V$ (Fig.~\ref{fig:HF}c), thus leading to four peaks in the DOS (Fig.~\ref{fig:HF}d). For $R_s\gg1$ the kinetic term is negligible and there is mixing between all four bands, again forming bonding and anti-bonding orbitals (Fig.~\ref{fig:HF}e). Since we can form bonding and anti-bonding orbitals in both the spin and the conduction/valence band degrees of freedom, this leads to a four-peak DOS (Fig.~\ref{fig:HF}f), where the splitting is set by $V$.

\textbf{Discussion ---} We used a combination of high-resolution STM and theoretical calculations to investigate the edge modes arising at a step edge on the surface of the topological insulator Pb$_{1-x}$Sn$_x$Se. We developed a continuum model description of these edge states and performed a Hartree-Fock calculation to investigate the effect of interactions. The edge modes have a flat dispersion, thus leading to ferromagnetic states, which may open up additional correlation gaps, as seen in the STM measurements on the system when doped to the Fermi level. In future work, it would be interesting to perform spin-resolved STM measurements on the edge modes to confirm that edge modes follow the symmetry breaking patterns predicted by the HF calculation. 

The step edge flat bands studied here have similarities to the edge states arising at the zig-zag edge of graphene. In graphene nanoribbons, the edges can be close enough such that they are coupled via interactions. In that case it is known that while the intra-edge coupling is ferromagnetic, the inter-edge coupling is antiferromagnetic. It is therefore natural to wonder what would happen with two nearby step edges in the TCI and how the edge modes are then coupled. Previous work has shown that two nearby step edge modes can couple to form bonding and anti-bonding orbitals \cite{PhysRevLett.126.236402}. It remains an open question what happens to the magnetism in that case.

\section{Acknowledgements}
GW would like to thank S. Parameswaran for useful discussions. TS, AS and JK thank P. Dziawa for SEM and R. Minikayev for XRD analysis. GW acknowledges NCCR MARVEL funding from the Swiss National Science Foundation. TS, AS and JK acknowledge the Foundation for Polish Science through IRA Programme co-financed by EU within Smart Growth Operational Programme for supporting crystal growth and characterization. We acknowledge support from the Deutsche Forschungsgemeinschaft (DFG, German Research Foundation) through QUAST FOR 5249-449872909 (Project P3). The work in W\"urzburg is further supported by the Deutsche Forschungsgemeinschaft (DFG, German Research Foundation) through Project-ID 258499086-SFB 1170 and the W\"urzburg-Dresden Cluster of Excellence on Complexity and Topology in Quantum Matter – ct.qmat Project-ID 390858490-EXC 2147. 

 	\bibliographystyle{unsrt}
 	\bibliography{bib.bib}

\clearpage 
\newpage

\onecolumngrid
	\begin{center}
		\textbf{\large --- Supplementary Material ---\\ Interaction effects in a 1D flat band at a topological crystalline step edge}\\
		\medskip
		\text{Glenn Wagner, Souvik Das, Johannes Jung,  Artem Odobesko,  Felix Küster,  Florian Keller,}\\  \text{Jedrzej Korczak,  Andrzej Szczerbakow,  Tomasz Story,  Stuart Parkin,} \\ \text{Ronny Thomale,  Titus Neupert, Matthias Bode,  and Paolo Sessi}
	\end{center}
	
		\setcounter{equation}{0}
	\setcounter{figure}{0}
	\setcounter{table}{0}
	\setcounter{page}{1}
	\makeatletter
	\renewcommand{\theequation}{S\arabic{equation}}
	\renewcommand{\thefigure}{S\arabic{figure}}
	\renewcommand{\bibnumfmt}[1]{[S#1]}
\begin{appendix}

\section{Supplementary figures}

\begin{figure}[h]
    \centering
    \includegraphics[width=0.75\textwidth]{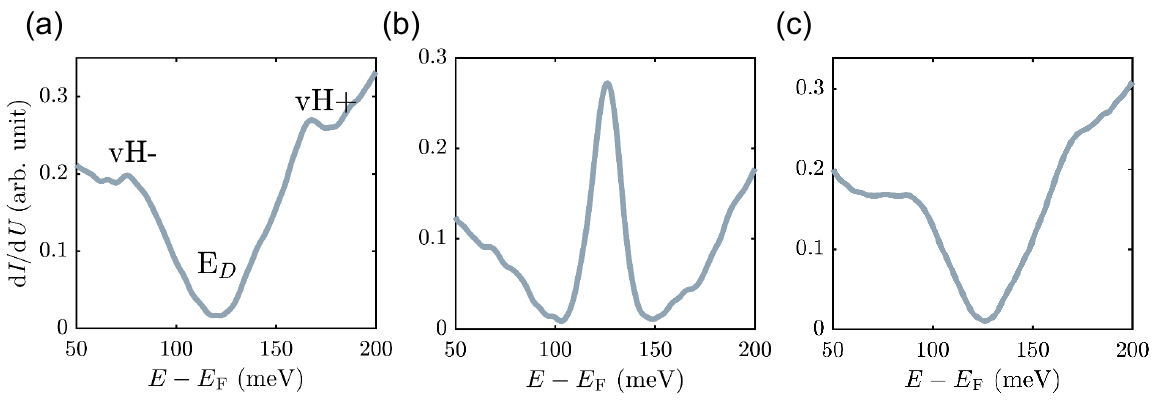}
    \caption{Scanning tunneling spectroscopy data acquired by positioning the tip on (a) terrace, (b) step corresponding to a structural $\pi$ shift, whose height corresponds to a half unit cell, and (c) step where translational invariance is preserved,corresponding to a unit cell height. vH- and vH+ identify the position of van Hove singularities energetically located at + 90 meV and + 175 meV,respectively. The minimum visible at $E_{\textrm{D}} \approx$ + 125 meV corresponds to the position of the Dirac point. At the energy corresponding to the Dirac point, a strong peak is visible in (b). As explained in Ref.~\onlinecite{Sessi}, this peak is the spectroscopic signature of 1D flat bands localized around the step.  }
    \label{fig:SuppSTS}
\end{figure}

\begin{figure}[h]
    \centering
    \includegraphics[width=0.45\textwidth]{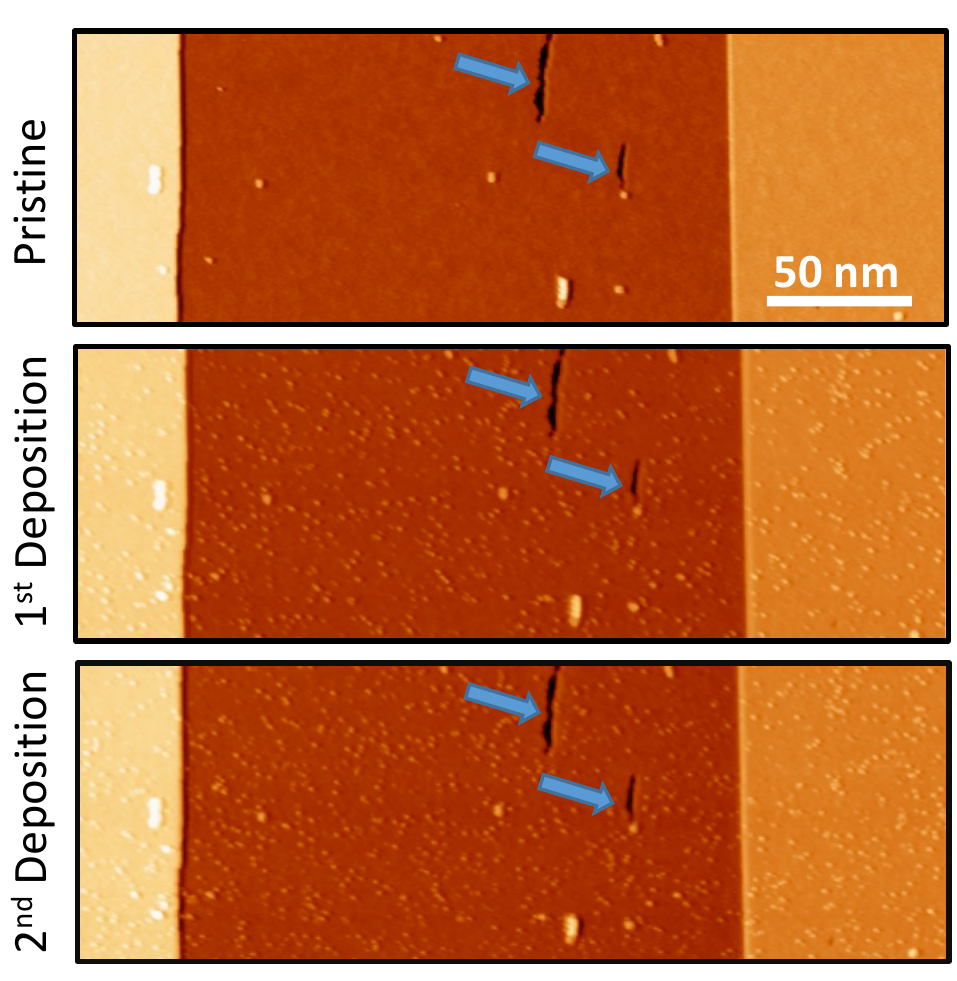}
    \caption{STM topographic images of the pristine Pb$_{1-x}$Sn$_x$Se surface (top panel) and after two deposition cycles (middle and bottom panels). Close inspection of the images acquired after deposition reveal small protrusions corresponding to Cr adatoms. The blue arrows point at some characteristic intrinsic defects, showing that images have been acquired on the very same sample region before and after deposition. This procedure ensure that the effects discussed in the main text are not related to sample inhomogeneities but that they are representative of the evolution of the electronic properties as a function of doping. }
    \label{fig:samelocation}
\end{figure}

\begin{figure}[h!]
    \centering
    \includegraphics[width=0.4\textwidth]{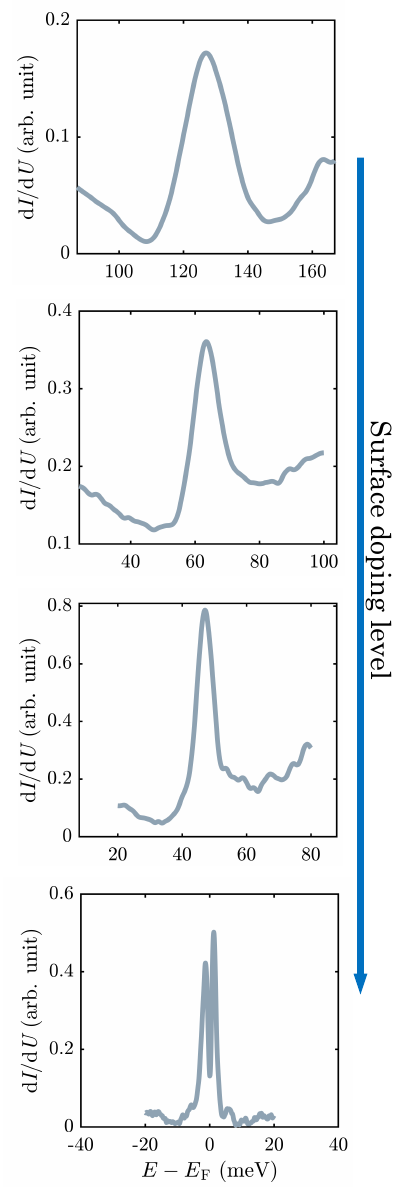}
    \caption{Doping dependent energy shift. (a-d) Scanning tunnelling spectroscopy of the flat band emerging at half unit cell steps as a function of the doping level. Cr adatoms deposited on the Pb$_{0.7}$Sn$_{0.3}$Se surface provide a $n$-doping effect. Starting from a $p$-doped crystal, this allows to progressively shift the energy of the 1D flat band towards the Fermi level.}
    \label{fig:SuppSplit}
\end{figure}

\begin{figure}[h!]
    \centering
    \includegraphics[width=0.75\textwidth]{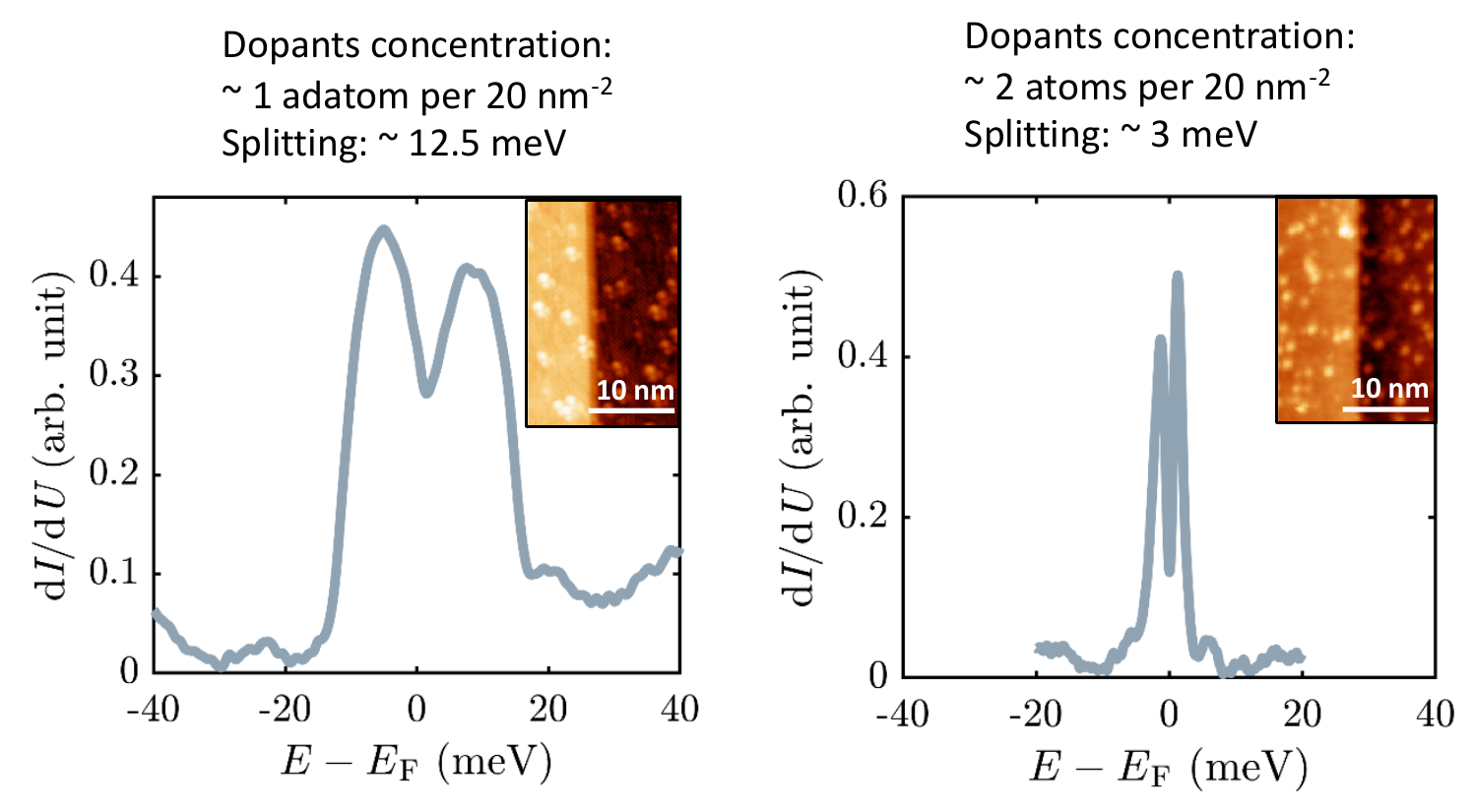}
    \caption{Comparison of the size of the splitting for samples characterized by different surface dopants concentration. The insets report topographic images corresponding to the areas where the respective spectroscopic data have been acquired. In both cases, Cr adatoms have been used as dopants.}
    \label{fig:Disorder}
\end{figure}

\begin{figure}[h]
    \centering
    \includegraphics[width=0.35\textwidth]{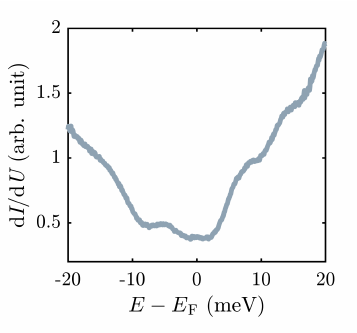}
    \caption{Scanning tunnelling spectroscopy acquired by positioning the tip on top of an integer step, i.e. a step with an height equals to the length of the unit cell. Surface dopants have been dosed to tune the Dirac point to the Fermi level. Contrary to spectra acquired on a structural $\pi$ shift, the spectrum is characterized by a minimum at the Dirac point and it shows a similar line shape compared to the pristine case. }
    \label{fig:integerstep}
\end{figure}

\newpage

\begin{figure}
    \centering
    \includegraphics[width=8cm]{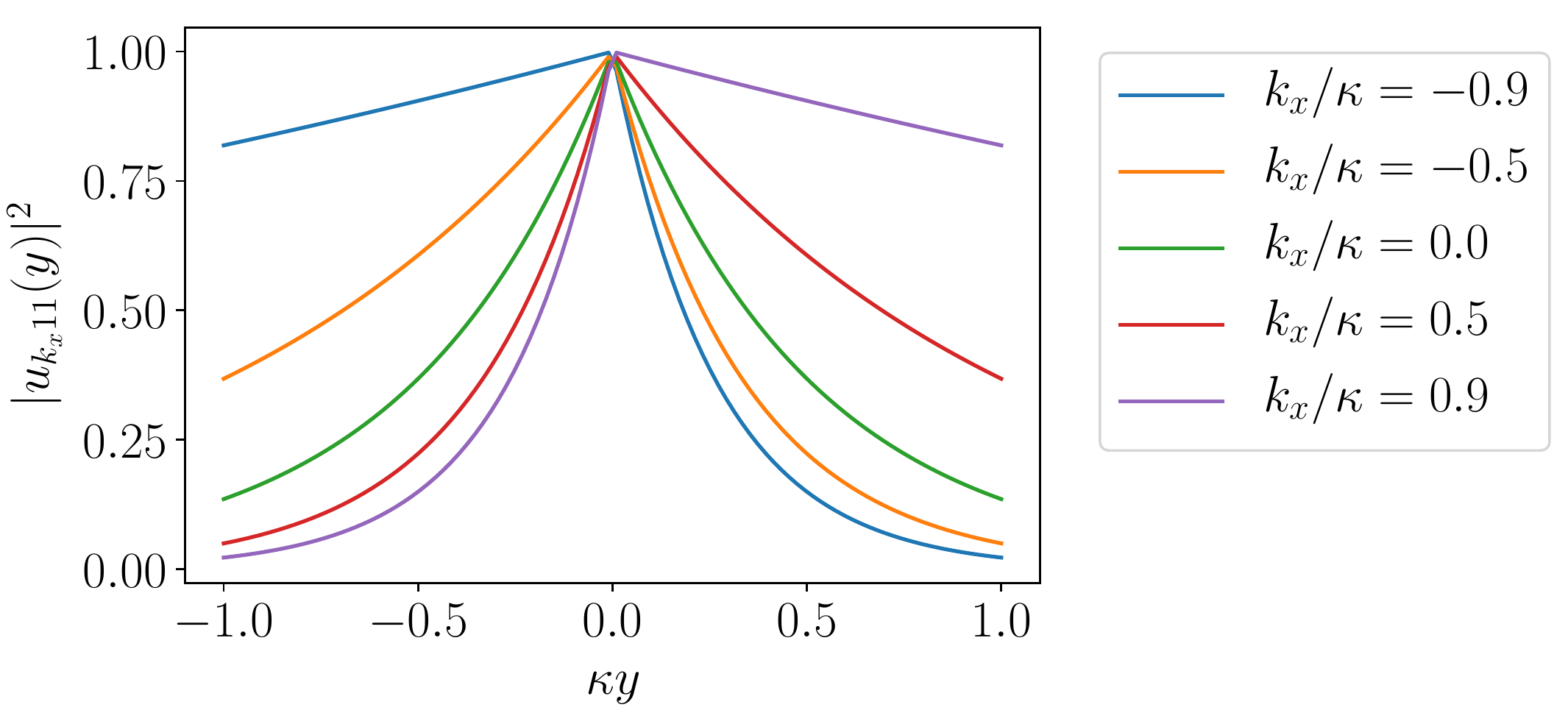} 
    \caption{The spatial profile of the wavefunctions in the direction perpendicular to the step edge for different momenta $k_x$.  The mean position $\langle y\rangle$ of the wavefunction shifts from left to right as $k_x$ increases. The maximum of the wavefunctions is always located at $y=0$.}%
    \label{fig:schemes}%
\end{figure}

\section{Hartree-Fock formalism}
From the main text we know that the edge states are strips akin to Landau gauge orbitals in the quantum Hall effect, which have the form
\begin{equation}
\psi_{k_x\tau\eta }(x,y)\equiv \frac{1}{\sqrt{L_x}}e^{ik_xx}u_{k_x\tau\eta}(y)=\frac{1}{\sqrt{L_x}}e^{ik_xx+i\kappa\eta y}\begin{cases}
e^{-|k_x-\tau\kappa|y}, \qquad y>0\\
e^{|k_x+\tau\kappa|y}, \quad\qquad y<0
\end{cases}
\label{eq:psi_zero_B_app}
\end{equation}
where $-\kappa<k_x<\kappa$. $L_x$ is the length of the system in the $x$-direction. We have time-reversal and mirror symmetries
\begin{align}
    \mathcal{T}&: \psi_{-k_x\bar\tau\bar\eta}(x,y)=\psi^*_{k_x\tau\eta }(x,y)\\
    \mathcal{M}_y&: \psi_{k_x\bar\tau\bar\eta}(x,y)=\psi_{k_x\tau\eta }(x,-y)\\
    \mathcal{M}_x&: \psi_{k_x\bar\tau\eta}(x,y)=\psi_{-k_x\tau\eta }(-x,y)\\
    \mathcal{P}&: \psi_{k_x\tau\bar\eta}(x,y)=\psi_{-k_x\tau\eta }(-x,-y)
\end{align}
Note that time-reversal flips both spin $\tau$ and valley $\eta$ and takes $k_x\to-k_x$. We want to project the Coulomb interaction down onto the Hilbert space of these edge states. The interaction term will be
\begin{equation}
   H_\textrm{int}= \frac{1}{2}V_{\alpha\beta\gamma\delta}c^\dagger_\alpha c^\dagger_\beta c_\delta c_\gamma,
\end{equation}
where we use the short-hand label $\alpha=(k_x^\alpha,\tau^\alpha,\eta^\alpha)$ and the matrix elements are
\begin{align}
    V_{\alpha\beta\gamma\delta}&=\langle\psi_\alpha\psi_\beta|\hat V|\psi_\delta\psi_\gamma\rangle\\
    &=\int_{\vec r,\vec r'}\psi_\alpha^*(\vec r)\psi_\beta^*(\vec r')V(\vec r-\vec r')\psi_\delta(\vec r)\psi_\gamma(\vec r')\\
    &=\frac{1}{L_x^2}\sum_{q_x}\int_{q_y}\int_{\vec r,\vec r'} e^{-ik_x^\alpha x-ik_x^\beta x'+ik_x^\delta x+ik_x^\gamma x'+iq_x(x-x')+iq_y(y-y')}V_q u_{k_x^\alpha\tau^\alpha\eta^\alpha}^*(y)u_{k_x^\beta\tau^\beta\eta^\beta}^*(y')u_{k_x^\delta\tau^\delta\eta^\delta}(y)u_{k_x^\gamma\tau^\gamma\eta^\gamma}(y'),
\end{align}
where $V_q=\frac{2\pi}{L_x}\frac{e^2}{2\epsilon_0q}$. We now define the (unnormalized) form factors
\begin{equation}
    \tilde\lambda_{\alpha\delta}(k_x^\alpha,k_x^\delta,q_y)=\int_y u_{k_x^\alpha\tau^\alpha\eta^\alpha}^*(y)u_{k_x^\delta\tau^\delta\eta^\delta}(y)e^{iq_yy}.
\end{equation}
The form factors can be calculated analytically by using the expression \eqref{eq:psi_zero_B_app} and one finds
\begin{equation}
   \tilde\lambda_{\alpha\delta}(k_x^\alpha,k_x^\delta,q_y)=\frac{i\kappa(\eta^\alpha-\eta^\delta)+iq_y+|k_x^\alpha-\tau^\alpha\kappa|+ |k_x^\delta-\tau^\delta\kappa|}{(\kappa(\eta^\alpha-\eta^\delta)+q_y)^2+(|k_x^\alpha-\tau^\alpha\kappa|+ |k_x^\delta-\tau^\delta\kappa|)^2}+\frac{-i\kappa(\eta^\alpha-\eta^\delta)-iq_y+|k_x^\alpha+\tau^\alpha\kappa|+ |k_x^\delta+\tau^\delta\kappa|}{(\kappa(\eta^\alpha-\eta^\delta)+q_y)^2+(|k_x^\alpha+\tau^\alpha\kappa|+ |k_x^\delta+\tau^\delta\kappa|)^2}.
\end{equation}
Note that the wavefunctions in \eqref{eq:psi_zero_B_app} are not normalized so the normalized form factors need to be calculated via
\begin{equation}
    \lambda_{\alpha\delta}(k_x^\alpha,k_x^\delta,q_y)=\frac{\tilde \lambda_{\alpha\delta}(k_x^\alpha,k_x^\delta,q_y)}{\sqrt{\tilde\lambda_{\alpha\alpha}(k_x^\alpha,k_x^\alpha,0)\tilde\lambda_{\delta\delta}(k_x^\delta,k_x^\delta,0)}}.
\end{equation}
The form factors have the form
\begin{equation}
    \lambda(k_x^\alpha,k_x^\delta,q_y)=(\Lambda^0(k_x^\alpha,k_x^\delta,q_y)\tau^0+\Lambda^0(-k_x^\alpha,k_x^\delta,q_y)\tau^x)\eta_0+(\Lambda^\times(k_x^\alpha,k_x^\delta,q_y)\tau^0+\Lambda^\times(-k_x^\alpha,k_x^\delta,q_y)\tau^x)\eta_x.
\end{equation}
Then the matrix elements become
\begin{align}
    V_{\alpha\beta\gamma\delta}&=\sum_{q_x}\int_{q_y} \delta_{q_x-k_x^\alpha+k_x^\delta,0}\delta_{-q_x-k_x^\beta+k_x^\gamma,0}\lambda_{\alpha\delta}(k_x^\alpha,k_x^\delta,q_y)\lambda_{\gamma\beta}^*(k_x^\gamma,k_x^\beta,q_y)V(q_x,q_y)\\
    &=\delta_{k_x^\alpha+k_x^\beta=k_x^\gamma+k_x^\delta}\int_{q_y} \lambda_{\alpha\delta}(k_x^\alpha,k_x^\delta,q_y)\lambda_{\gamma\beta}^*(k_x^\gamma,k_x^\beta,q_y)V(k_x^\alpha-k_x^\delta,q_y),
\end{align}
where the Kronecker delta in the last line is enforcing momentum conservation in the $x$-direction.
Now we perform the mean-field decoupling of the Hamiltonian
\begin{equation}
   H_\textrm{int,HF}= V_{\alpha\beta\gamma\delta}(c^\dagger_\alpha c_\delta\langle c^\dagger_\beta  c_\gamma\rangle-c^\dagger_\alpha c_\gamma\langle c^\dagger_\beta c_\delta \rangle).
\end{equation}
A Slater determinant is described by the projector $P_{\alpha\beta}=\langle c^\dagger_\alpha c_\beta \rangle$. Assuming translational invariance along the $x$-direction, we have 
\begin{equation}
P_{\alpha\beta}(k_x^\alpha,k_x^\beta)=\langle c^\dagger_\alpha(k_x^\alpha) c_\beta (k_x^\beta)\rangle=\delta_{k_x^\alpha,k_x^\beta}\langle c^\dagger_\alpha(k_x^\alpha) c_\beta (k_x^\alpha)\rangle\equiv\delta_{k_x^\alpha,k_x^\beta} P_{\alpha\beta}(k_x^\alpha).
\end{equation}
With this simplification the HF Hamiltonian becomes
\begin{align}
   H_\textrm{int,HF}&= V_{\alpha\beta\gamma\delta}(c^\dagger_\alpha(k_x^\alpha) c_\delta(k_x^\alpha) P_{\beta \gamma}(k_x^\beta)\delta_{k_x^\beta,k_x^\gamma}-c^\dagger_\alpha(k_x^\alpha) c_\gamma(k_x^\alpha) P_{\beta\delta}(k_x^\beta)\delta_{k_x^\beta,k_x^\delta})\\
   &=V_{\alpha\beta\gamma\delta}^D(k_x^\alpha,k_x^\beta)P_{\beta \gamma}(k_x^\beta)c^\dagger_\alpha(k_x^\alpha) c_\delta(k_x^\alpha)-V_{\alpha\beta\gamma\delta}^E(k_x^\alpha,k_x^\beta)P_{\beta \delta}(k_x^\beta)c^\dagger_\alpha(k_x^\alpha) c_\gamma(k_x^\alpha),
\end{align}
where we used the momentum-conserving Kronecker delta and we defined the direct and exchange matrix elements
\begin{align}
    V_{\alpha\beta\gamma\delta}^D(k_x^\alpha,k_x^\beta)&=\int_{q_y} \lambda_{\alpha\delta}(k_x^\alpha,k_x^\alpha,q_y)\lambda_{\gamma\beta}^*(k_x^\beta,k_x^\beta,q_y)V(0,q_y),\\
    V_{\alpha\beta\gamma\delta}^E(k_x^\alpha,k_x^\beta)&=\int_{q_y} \lambda_{\alpha\delta}(k_x^\alpha,k_x^\beta,q_y)\lambda_{\gamma\beta}^*(k_x^\alpha,k_x^\beta,q_y)V(k_x^\alpha-k_x^\beta,q_y).
\end{align}

To this interaction Hamiltonian we add a kinetic component
\begin{equation}
    H_\textrm{kin}=\epsilon_{\alpha}(k_x)c_\alpha^\dagger(k_x)c_\alpha(k_x)
\end{equation}
with the phenomenological bandstructure (chosen to reproduce the key features of the bandstructure found in \cite{Sessi})
\begin{equation}
    \epsilon_{\tau\eta}(k_x)=W\eta\bigg(\tau\cos\frac{\pi k_x}{2\kappa}+\frac{1}{5}\sin\frac{\pi k_x}{\kappa}\bigg)
\end{equation}
with $W$ the bandwidth. We now perform self-consistent Hartree-Fock calculations on the Hamiltonian
\begin{equation}
    H_\textrm{HF}=H_\textrm{kin}+H_\textrm{int,HF}.
\end{equation}

 	\end{appendix}
 	\end{document}